\documentclass[10pt]{article}
\makeatletter
\usepackage{multicol}
\input epsf
\begin{document}

\title{A model for interacting instabilities and texture dynamics of patterns}
\author{ Alaka Das, Krishna Kumar and Narayanasamy Ganesh\\
{\small {\it Physics and Applied Mathematics Unit, Indian
Statistical Institute, 203, B.~T.~Road,  Calcutta-700~035, India }}\\
(\today)
\date{}}
\maketitle

\begin{center}
{\small{
\begin{minipage}[t]{14cm}
\hspace*{0.5cm} A simple model to study interacting
instabilities and textures of resulting patterns for thermal
convection is presented. The model consisting of twelve-mode
dynamical system derived for periodic square lattice describes
convective patterns in the form of stripes and patchwork quilt.
The interaction between stationary zig-zag stripes  and standing
patchwork quilt pattern  leads to spatiotemporal patterns of {\it
twisted} patchwork quilt. Textures of these patterns, which
depend strongly on  Prandtl number, are investigated numerically
using the model. The model also shows an interesting possibility
of a multicritical point, where stability boundaries of four
different structures meet.\\ \\
\noindent PACS number(s): 47.54.+r, 47.20.Lz, 05.45.Ac
\end{minipage}
}}
\end{center}

\begin{multicols}{2}
Pattern-forming instabilities in systems driven externally far
from equilibrium are currently receiving considerable attention
[1-11]. They  appear in many physical systems such as fluids
[3-8], granular materials [9], cardiac tissues [10],
reaction-diffusion systems [11], traffic flow [12], dendritic
growth [13], and non-linear optics [14]. Spatiotemporal structures
arising due to interacting instabilities and the dynamics of
their textures  are understood theoretically either by amplitude
equations [1-2] or dynamical systems [15]. In absence of a clear
separation of time scales, dynamical systems are preferred for
investigating pattern dynamics. Roberts {\it et al} [16], using a
dynamical system with hexagonal symmetry, showed the possibility
of standing patchwork quilt pattern due to interacting
oscillatory instabilities in the problem of thermal convection in
a {\it double-diffusive} system [17]. They found mirror-symmetric
and {\it twisted} patchwork quilt patterns on hexagonal lattice.
Patterns having both open and closed streamlines are called
patchwork quilt. {\it Twisted} patchwork quilt does not have
mirror symmetry.

In this work, we present a simple model of interacting
instabilities  in the form of a twelve-mode dynamical system
derived from Boussinesq equations for thermal convection in {\it
ordinary} fluids. Using the model, we show that the interaction
between  zig-zag  pattern and standing squares can also lead to
{\it twisted} patchwork quilt pattern. Our model is based on
square lattice rather than hexagonal lattice, and it requires
only two bifurcation parameters: the Prandtl number $\sigma $ and
reduced Rayleigh number $r$. The possibility of patchwork quilt
patterns on square lattice due to interaction of a stationary and
an oscillatory instabilities is qualitatively new. We then
investigate numerically textures of spatiotemporal structures
arising due to interacting patterns. The model also shows an
interesting possibility of a multicritical point ($\sigma =1.57
\pm 0.01,\ r=11.2 \pm 0.05$), where stability zones of straight
stripes, zig-zag stripes, standing  {\it symmetric} patchwork
quilt, and standing {\it twisted} patchwork quilt meet.


We consider an extended horizontal layer of Boussinesq fluid of
thickness $d$, kinematic viscosity $\nu$, thermal diffusitivity
$\kappa$ confined between two  perfectly conducting {\it
stress-free} horizontal boundaries, and heated from below. Making
all length scales dimensionless by the fluid thickness $d$, time
by the thermal diffusive time scale $d^2/\kappa$, and the
temperature  by the temperature difference $\Delta T$ between the
two bounding surfaces, the relevant hydrodynamical equations in
dimensionless form  read
\begin{eqnarray}
\partial_t{\nabla}^2v_3 &=& \sigma{\nabla}^4v_3
+\sigma{{\nabla}^2_H}\theta \nonumber \\
&-& {\bf{e_3}}.[{\bf{\nabla}}\times \{ ({\bf{\omega}}.{\bf
{\nabla}}){\bf v} - ({\bf v}.{\bf{\nabla}}){\bf{\omega}}\} ],\\
\partial_t{\omega_3} &=& \sigma{\nabla}^2{\omega_3} + [({\bf{\omega}}.{\bf
 {\nabla}}){v_3}-({\bf v}.{\bf{\nabla}}){\omega_3}],\\
 \partial_t\theta &=& {\nabla}^2\theta+ R v_3 - {\bf
 v}.{\bf{\nabla}}\theta,
\end{eqnarray}
 where ${\bf v} \equiv (v_1,v_2,v_3)$, $\omega =$${\nabla}\times{\bf v}$
 $\equiv$ $(\omega_1,\omega_2, \omega_3)$, and $\theta$ are respectively
 the velocity, the vorticity, and the deviation from the conductive
 temperature profile. Prandtl number $\sigma$ and Rayleigh number $R$ are
 defined respectively as $\sigma = \frac{\nu}{\kappa}$ and
 $R = \frac{\alpha(\Delta T)g{d}^3}{\nu\kappa}$, where $\alpha$ is the
 coefficient of thermal expansion of the fluid, $g$ the acceleration due to
 gravity. The unit vector ${\bf e_3}$ is directed vertically upward. The
 symbol $\nabla_H^2 (= \nabla_{11} + \nabla_{22})$ stands for horizontal
 Laplacian. The boundary conditions at the idealized {\it stress-free}
 conducting flat surfaces imply
 $\theta$ $ = $ $ v_3$ $ = $ $\partial_{33}{v_3} $ $ = $ $\partial_{3}{\omega_3}$
 $ = 0$ at $x_3 = 0, 1$.

 We construct a dynamical system by standard Galerkin procedure.
 The spatial dependence of vertical velocity, vertical vorticity and
 temperature field are expanded in a Fourier series, which is compatible
 with the stress-free flat conducting boundaries and periodic square lattice
 in the horizontal plane. We include minimum modes to describe straight
 stripes (S), zig-zag stripes (ZZ), square patterns (SQ), and nonlinear
 interaction among these patterns. The vertical velocity $v_3$,
 vertical vorticity $w_3$, and $\theta$ then may be written as
\begin{eqnarray} v_3 &=& [W_{101}(t) \cos{k_cx_1}  + W_{011}(t)
\cos{k_cx_2}]\sin{\pi x_3}
\nonumber \\
 &+& [W_{112}(t) \cos{k_cx_1}\cos{k_cx_2} \nonumber \\
 &&~~~~+~ W_{\bar{1}\bar{1}2} \sin{k_cx_1}\sin{k_cx_2}] \sin{2{\pi}x_3},\\
\omega_3 &=& [\zeta_{101}(t) \cos{k_cx_1} + \zeta_{011}(t) \cos{k_cx_2}]
 \cos{\pi x_3} \nonumber \\
  &+& \zeta_{110}(t) \cos{k_cx_1}\cos{k_cx_2},\\
\theta &=& [\Theta_{101}(t)\cos{k_cx_1} +
\Theta_{011}(t)\cos{k_cx_2}]\sin{{\pi}x_3} \nonumber \\
&+& \Theta_{002}(t)\sin{2{\pi}{x_3}} +[\Theta_{112}(t)\cos{k_cx_1}\cos{k_cx_2} \nonumber \\
&&~~~~+~ \Theta_{\bar{1}\bar{1}2}\sin{k_cx_1}\sin{k_cx_2}]
\sin{2{\pi}x_3},\end{eqnarray} where $k_c = \pi/\sqrt{2} $. The
horizontal components of velocity and vorticity fields are
computed by the solenoidal characters of these two fields (i.e.,
$\nabla \cdot {\bf v} = \nabla \cdot {\bf \omega} = 0$). We now
project the hydrodynamical equations ($1-3$) onto these twelve
modes to get the following dynamical system
\end{multicols}
\begin{eqnarray}
\tau\dot{\bf{X}} &=& \sigma(-{\bf{X}}+{\bf{Y}})+
\left(\begin{array}{l} X_{2} \\  X_{1} \end{array} \right) S_{1}
+ \left(\begin{array}{l} ~~G_{2} \\ -G_{1}  \end{array}
\right)S_{2} -  \left(\begin{array}{l} G_{2} \\  G_{1}
\end{array} \right)
V,\label{modelbegin}  \\
\tau\dot{\bf{Y}} &=& -{\bf{Y}}+(r-Z) {\bf{X}} +
\left(\begin{array}{l}  X_{2}\\ X_{1} \end{array} \right)
 T_{1}
+ \left(\begin{array}{l} ~~G_{2} \\  -G_{1} \end{array} \right)
 T_{2},  \\
\tau\dot{\bf{G}} &=& -\sigma {\bf G} + \frac{2}{3}
 \left(\begin{array}{l} ~~X_{2} \\ -X_{1}   \end{array} \right)  S_{2}
 + \left(\begin{array}{l}  G_{2} \\  G_{1}    \end{array} \right)  S_{1},\\
\tau\dot{\bf{S}} &=& - \frac{10}{3} \sigma {\bf S}+ \frac{3}{5}
\sigma {\bf{T}} -
 \left(\begin{array}{l} \frac{3}{10}(X_{1}X_{2}+ G_{1}G_{2}) \\
 \frac{3}{40}(X_{1}G_{2}-X_{2}G_{1}) \end{array} \right),\\
\tau\dot{\bf{T}} &=& -\frac{10}{3}{\bf{T}}+r{\bf{S}} -
 \left(\begin{array}{l} \frac{1}{4}(X_{1}Y_{2} + X_{2}Y_{1}) \\
 \frac{3}{8}(Y_{1}G_{2} - Y_{2}G_{1})\end{array} \right),\\
\tau\dot{V} &=& -\frac{2}{3} \sigma V+(X_{1}G_{2} + X_{2}G_{1}),\\
\tau\dot{Z} &=& - \frac{8}{3} Z + (X_1 Y_1 + X_2 Y_2),
\label{modelend}
\end{eqnarray}
\begin{multicols}{2}
\noindent where the critical modes ${\bf X}$ $\equiv$
$(X_1,~X_2)^T$ $ = $ $\frac{\pi}{\sqrt{2} q_c^2}$
$(W_{101},~W_{011})^T$, ${\bf Y}$ $\equiv$ $(Y_1,~Y_2)^T$ $=$
$\frac{\pi {k_c}^2}{\sqrt{2}q_c^6}$
$(\Theta_{101},~\Theta_{011})^T$, and ${\bf G}$ $\equiv$
$(G_1,~G_2)^T$ $=$  $\frac{\pi}{\sqrt{2} q_c^3}$
$(\zeta_{101},~\zeta_{011})^T$ are proportional to vertical
velocity, temperature, and vertical vorticity respectively. The
non-linear modes are redefined as ${\bf S}$ $\equiv$
$(S_1,~S_2)^T$ $=$ $ \frac{1}{4q_c}$
$(\frac{\pi}{q_c}W_{112},~W_{\bar{1}\bar{1}2})^T$, $V$ $=$
$\frac{\pi}{2q_c^3}$$\zeta_{110} $, ${\bf T}$ $\equiv$
$(T_1,~T_2)^T$ $=$ $\frac{k_c^2 }{4q_c^5}$ $(\frac{\pi}{q_c}
\Theta_{112},~\Theta_{\bar{1}\bar{1}2})^T,$ and $Z$ $=$
$-\frac{\pi k_c^2 }{q_c^6}$ $\Theta_{002}.$ The constants of the
model are $q_c^2= \pi^2 + k_c^2$ and $\tau = q_c^{-2}.$ Prandtl
number $\sigma $ and the reduced Rayleigh number $r =
{\frac{R}{R_c}} \left(= {\frac{Rk_c^2}{q_c^6}} \right)$ are two
bifurcation parameters of our model. the superscript $T$ denotes
the transpose of a matrix.

The model (\ref{modelbegin} - \ref{modelend}) describes various
stationary as well as oscillating patterns on square lattice. The
set of straight stripes (S) parallel to $x_{1(2)}$-axis is
obtained by setting $X_{2(1)}$ $=$ $Y_{2(1)}$ $=$ $G_1$ $=$ $G_2$
$=$ $S_1$ $=$ $S_2$ $=$ $T_1$ $=$ $T_2$ $=$ $V$ $=$ $0$ in the
model. The stationary straight stripes given by $X_{1(2)}$ $=$
$Y_{1(2)}$ $=$ $\sqrt{8(r-1)/3},$ and $Z=r-1$ appear just above
onset ($r=1$) of convective instability. The stationary zig-zag
(ZZ) patterns, which appear at secondary instability  for
$(\sigma < 1.57),$ are obtained by taking $X_{2(1)}$ $=$
$Y_{2(1)}$ $=$ $G_{1(2)}$ $=$ $S_{1(2)}$ $=$ $T_{1(2)}$ $=0$ in
the model. The standing asymmetric squares~\cite{das} is
retrieved by setting $G_1$ $=$ $G_2$ $=$ $S_2$ $=$ $T_2$ $=$ $V$
$=$ $0$ in the model. The asymmetric squares, which form
mirror-symmetric patchwork quilt pattern, appear at the onset of
secondary instability via forward Hopf bifurcation for $\sigma >
1.57 $. The twelve-mode model describe interaction among these
structures. We integrate numerically the full model to
investigate dynamics of the resulting convective structures. We
do it for a fixed $\sigma$ by varying $r$ in small steps. For
each value of $r$, the integration is done starting with randomly
chosen initial conditions  for long enough to reach the final
state. Prandtl number $\sigma $ is then varied in small steps and
whole procedure is repeated for each $\sigma$. The final states
for various $\sigma $ and $r$ reported here are independent of
the choice of initial conditions.

 Figure 1 shows the stability boundaries of various
patterns in parameter space ($\sigma-r$ plane) computed from the
model. A transition from straight stripes (S) to zig-zag stripes
(ZZ) occur as $r$ is raised above its value at the lower
stability boundary for $\sigma < 1.57$. The threshold value of
$r$ for such transition strongly depends on $\sigma$.  The
transition from straight stripe to standing patchwork  quilt (PQ)
via forward Hopf bifurcation occurs when $r$ is raised above its
value at the lower boundary for $\sigma > 1.57$. The patchwork
quilt pattern shows mirror and inversion symmetries but not
four-fold symmetry. A shadowgraph of this pattern appears as
standing asymmetric squares [18]. The stability boundary of this
Hopf bifurcation shows weak dependence on $\sigma$.

All of the straight stripes (S), zig-zag stripes (ZZ) and standing
patchwork quilt (PQ ) are unstable in the region of parameter
space marked as TPQ. We find all modes of the model active, if
$\sigma$ and $r$ are chosen from this zone, and interacting with
each other. We observe spatiotemporal patterns without mirror
symmetry in this part of parameter space. Figure 2 shows the {\it
twisted} patchwork quilt pattern slightly above the multicritical
point.  These patterns have lost the mirror symmetry. This
happens due to competition of asymmetric squares, which are mirror
symmetric patchwork quilt pattern, with zig-zag patterns. The
generation of vertical vorticity breaks the mirror symmetry of
patchwork quilt pattern (PQ) as $\sigma$ and $r$ are chosen from
the zone marked TPQ  in parameter space (see Fig.1). An increase
in the intensity of vertical vorticity makes the pattern  more
twisted. The set of four figures clearly depicts the
spatio-temporal behaviour of the texture of the {\it twisted}
patchwork quilt patterns. The texture depends strongly on $\sigma
$ and weakly on $r$. Figure 3 shows competition of two sets,
mutually perpendicular to each other, competing with each other.
The picture for $\sigma=0.6$ and $r=9.0$ shows periodically
varying textures arising due to competing instabilities for one
period of oscillation. The model also shows chaotic patterns for
$\sigma = 0.835$ and $r=11.4$ (see fig. 4). This chaotic
evolution of patterns occurs via quasi-periodic route.

In this article, we have presented a simple model  of interacting
instabilities. We have shown that the interaction between a
stationary instability and an oscillatory instability may lead to
many interesting patterns including {\it twisted} patchwork quilt
on square lattice. The texture of the patterns due to competing
instabilities may be modeled with an appropriate dynamical
system. The model is also useful in  capturing mechanism of
emergence of various instabilities and resulting patterns.

We acknowledge support from DST, India through its grants under
the project ``Pattern-forming instabilities and interface waves".
N Ganesh of I I T, Bombay acknowledges partial support from PAMU,
ISI, Calcutta.

\small{

}
\end{multicols}

\twocolumn
\begin{figure}
\centerline{\epsfxsize=2.5in \epsfbox{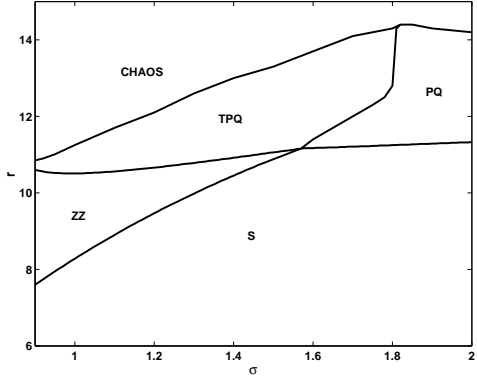}} \caption{{\small
Stability boundaries of various convective structures in
parameter space computed by the model. Stability zones of
straight stripes (S), zig-zag stripes (ZZ), patchwork quilt (PQ),
and {\it twisted} patchwork quilt (TPQ) meet at  a multicritical
point ($\sigma=1.57 \pm 0.01,\ r=11.2 \pm 0.05$). The model shows
chaotic behaviour at much higher values of $r$.}} \label{fig1}
\end{figure}

\begin{figure}
\centerline{\epsfxsize=3.0in\epsfbox{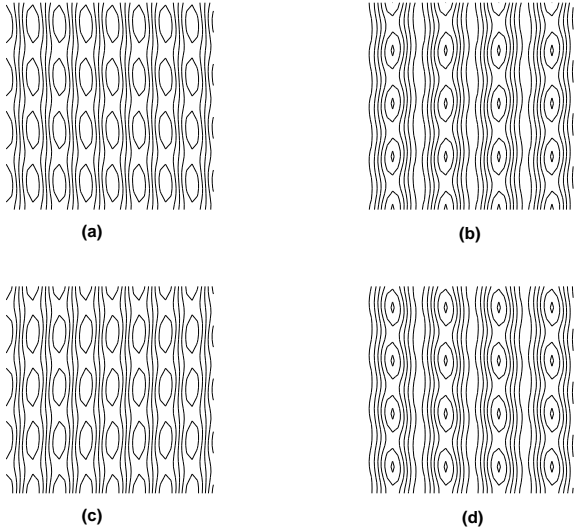}} \caption{{\small
Texture of {\it twisted} patchwork quilt patterns. Stream lines
for $\sigma=1.57,\ r=11.4, z=0.25$ at (a) $t=0$, (b) $t=T/4$, (c)
$t=T/2$, and (d) $t=3T/4$.}} \label{fig2}
\end{figure}

\begin{figure}
\centerline{\epsfxsize=3.0in\epsfbox{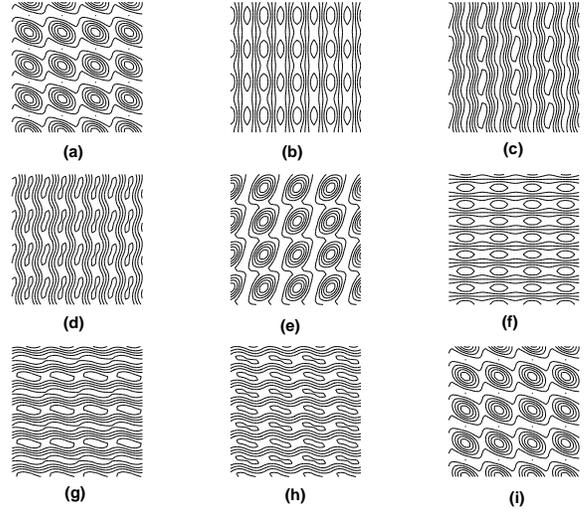}} \caption{{\small
Temporal sequence of textures of structures due to competition of
standing mirror-symmetric and {\it twisted patchwork quilt}
patterns for $\sigma=0.6,\ r=9.0, z=0.25$ at (a) $t=0$, (b)
$t=T/8$, (c) $t=T/4$, (d) $t=3T/8$, (e) $t=T/2$, and (f)
$t=5T/8$, (g) $t=3T/4$, (h) $t=7T/8$, (i) $t=T$.}} \label{fig3}
\end{figure}

\begin{figure}
\centerline{\epsfxsize=3.0 in\epsfbox{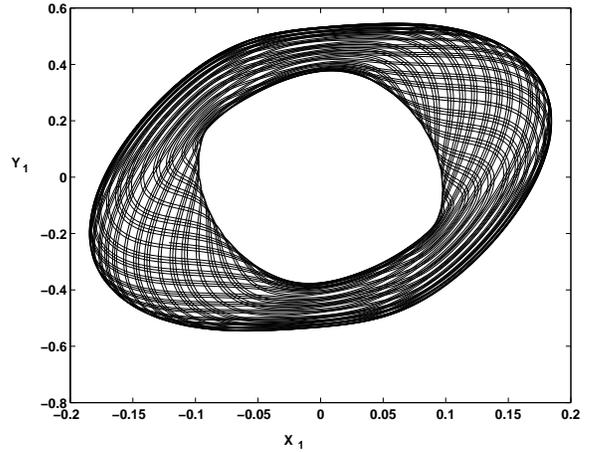}} \caption{{\small
Chaotic patterns for $\sigma=0.835,\ r=11.4$. The chaos occurs
via quasi-periodicity.}}\label{fig4}
\end{figure}

\end{document}